\pdfoutput=1

\documentclass[11pt,table]{article}

\usepackage[preprint]{acl}
\usepackage[ruled,vlined]{algorithm2e}
\usepackage{times}
\usepackage{latexsym}
\usepackage{multirow}
\usepackage{cite}
\usepackage{xcolor}
\usepackage{makecell}

\usepackage[normalem]{ulem}
\usepackage{amsfonts,amsmath,amssymb}
\usepackage{amsthm}
\usepackage{booktabs}
\usepackage{lscape}
\usepackage{breqn}
\usepackage{pifont}
\usepackage[T1]{fontenc}

\NewDocumentCommand\emojismiley{}{
    \includegraphics[scale=0.08]{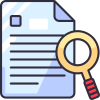}
}

\usepackage[utf8]{inputenc}

\usepackage{microtype}

\usepackage{inconsolata}

\usepackage{graphicx}

\usepackage{bm}
\newcommand{\EQ}{\begin{eqnarray}}
\newcommand{\EN}{\end{eqnarray}}

\usepackage{tabulary}

\newcommand*{\rowstyle}[1]{
	\gdef\@rowstyle{\leavevmode#1}%
	\@rowstyle\ignorespaces}

\usepackage{multicol}
\usepackage{algorithmic}
\title{AnalyticKWS: Towards Exemplar-Free\\ Analytic Class Incremental Learning for Small-footprint Keyword Spotting}

\author{
 \textbf{Yang Xiao\textsuperscript{1}\textsuperscript{2}},
 \textbf{Tianyi Peng\textsuperscript{3}},
 \textbf{Rohan Kumar Das\textsuperscript{2}},
 \textbf{Yuchen Hu\textsuperscript{3}},
 \textbf{Huiping Zhuang\textsuperscript{4}},
\\
 \textsuperscript{1}The University of Melbourne
 \textsuperscript{2}Fortemedia Singapore
 \textsuperscript{3}Nanyang Technological University \\
 \textsuperscript{4}South China University of Technology
\\
 \textsuperscript{}{\small{\textit {yxiao9550@student.unimelb.edu.au}}}
}

\newcommand{\cmark}{\ding{51}}%
\newcommand{\xmark}{\ding{55}}%
\begin{document}
\maketitle
\begin{abstract}
Keyword spotting (KWS) offers a vital mechanism to identify spoken commands in voice-enabled systems, where user demands often shift, requiring models to learn new keywords continually over time. However, a major problem is catastrophic forgetting, where models lose their ability to recognize earlier keywords. Although several continual learning methods have proven their usefulness for reducing forgetting, most existing approaches depend on storing and revisiting old data to combat catastrophic forgetting. Though effective, these methods face two practical challenges: 1) privacy risks from keeping user data and 2) large memory and time consumption that limit deployment on small devices. To address these issues, we propose an exemplar-free Analytic Continual Learning (AnalyticKWS~\emojismiley) method that updates model parameters without revisiting earlier data. Inspired by efficient learning principles, AnalyticKWS computes a closed-form analytical solution for model updates and requires only a single epoch of adaptation for incoming keywords.  AnalyticKWS demands fewer computational resources by avoiding gradient-based updates and does not store old data. By eliminating the need for back-propagation during incremental learning, the model remains lightweight and efficient. As a result, AnalyticKWS meets the challenges mentioned earlier and suits resource-limited settings well. Extensive experiments on various datasets and settings show that AnalyticKWS consistently outperforms existing continual learning methods. 
\end{abstract}

\section{Introduction}

As a key component of edge intelligence, devices such as robots, autonomous systems, and smart assistants interact naturally with humans through voice~\citep{smarthome,zhang2018hello,mandal2014recent}. Spoken keyword spotting (KWS)~\citep{lopezespejo2021deep} identifies specific keyword phrases within recorded speech and is essential for edge computing devices. These devices require quick responses, low energy consumption, and high accuracy to meet user demands. Cloud-based solutions may not be ideal in these setups because sending private data to a remote server can violate privacy rules, and real-time updates often require immediate on-device adaptation. Due to the KWS system always being applied in practical real-world scenarios, modern small-footprint KWS systems~\citep{tang2018kws,choi2019temporal,kim2021broadcasted,ng23b_interspeech} based on deep learning often use compact models to balance performance and computational cost. However, these systems face significant challenges as their performance usually drops when encountering new keywords in the target domain. 

With the increasing demand for voice as the mode for interaction-oriented tasks in embodied AI, it is important to support more personalized applications~\citep{yang2022personalized}, such as smart home devices and in-car assistants. These devices must continuously learn new keywords while respecting user privacy and resource limits. However, re-training a KWS model from scratch with new keywords is not only time-consuming, but also resource-intensive. Previous work~\citep{awasthi2021teaching,2021Few, parnami2022few} addresses this issue through a few-shot fine-tuning, which adapts a model to target data with minimal samples but suffers from catastrophic forgetting~\citep{mccloskey1989catastrophic}, where previous knowledge deteriorates.

To solve the forgetting issue, continual learning (CL)~\citep{cl} integrates new data while retaining previous knowledge. Within this framework, class incremental learning (CIL) \citep{cil1} focuses on adding new classes to the model sequentially, making it especially relevant for KWS that involve evolving label sets. Recent CIL strategies split into rehearsal-based methods that store past examples for future training and exemplar-free approaches that do not keep old data. For rehearsal-based methods, Xiao et al.~\citep{cl2} first suggested choosing examples through a diversity-based approach for KWS. Based on that, the latest works of Peng et al.~\citep{peng2024dark} further saved model predictions to distill prior knowledge. However, due to the constraints present in real-world, rehearsal-based CIL is often not reliable for KWS. First, storing past examples risks breaching user privacy. Second, it consumes excessive memory, which is not feasible for resource-limited edge devices.


Although exemplar-free class incremental learning (EFCIL) methods avoid storing historical data and thus bypass privacy concerns~\citep{goswami2024resurrecting, zhuang2022acil,szatkowski2024adapt,pclkws2022}, many of these methods still rely on complex optimizers or dynamic network structures. This approach can be unsuitable for edge devices, which lack the computational power for extensive gradient-based updates. Hence, we propose a more efficient method that preserves the benefits of EFCIL but removes the need for complex adaptations, making it more practical for resource-constrained KWS systems.

As we mentioned, the key challenge in incremental KWS is catastrophic forgetting, where new keywords overwrite knowledge of previous ones. Existing solutions address this issue but often store prior data, creating privacy risks and high memory use. To avoid these concerns, we propose AnalyticKWS~\emojismiley, an exemplar-free method that mitigates catastrophic forgetting while eliminating the need for using past examples. Drawing on analytic learning~\citep{brmp}, AnalyticKWS uses a recursive least-squares procedure in place of back-propagation, letting it incorporate new knowledge and protect user data. Our core contribution is to maintain previous knowledge without retaining retrospective data, thus resolving catastrophic forgetting in a privacy-preserving and resource-efficient manner. We evaluate the proposed AnalyticKWS for a wide range of incremental KWS task settings to demonstrate its effectiveness. Moreover, by processing new keywords in a single forward pass without gradient updates, AnalyticKWS has the capability to lower the computational overhead making it ideal for edge devices. The primary contributions of this paper can be summarized as follows:


\begin{itemize}

\item \textbf{Mitigate Forgetting:} AnalyticKWS reduces catastrophic forgetting by preserving the knowledge of past tasks without using historical data. Comprehensive experiments on three datasets with up to 100 keywords are conducted to compare AnalyticKWS with other baselines to project its effectiveness for incremental KWS. 
\vspace{-1mm}
    \item \textbf{Privacy and Memory Efficiency:} We propose AnalyticKWS, which adopts a frozen acoustic feature extractor and an analytic classifier without retaining  any past data. By eliminating exemplars, this design enhances user privacy and reduces memory usage, making it suitable for devices with limited resources.
\vspace{-1mm}
\item \textbf{Low Computational Overhead:} During CL, our method updates the analytic classifier in a single step without requiring gradient back-propagation. We measure both training time and extra memory to project the capability of AnalyticKWS with fewer resources and adaptation to new keywords within a single epoch, meeting the demands of real-world, resource-constrained environments.
\end{itemize}

\section{Related Work}
\textbf{Small-footprint Keyword Spotting:} 
With the widespread adoption of voice interfaces in smart consumer electronics, the application of small convolutional neural networks in compact KWS has become increasingly significant. Recent works investigated innovative convolution techniques to improve KWS performance. Chen et al.~\citep{chen2014small} were the first to apply deep neural networks to treat KWS as a classification task. TC-ResNet proposed in~\citep{choi2019temporal} applies 1D temporal convolution to enhance efficiency and accuracy. The authors of~\citep{kim2021broadcasted} introduced broadcasted residual learning in BC-ResNet combining 1D and 2D convolutions. Despite the effectiveness, these methods are typically trained with a limited set to reduce computation and memory usage. However, users need to customize a new set of voice commands to suit their environment. In this work, we investigate the CL to develop a dynamic KWS approach while incrementally learning from unseen keywords. \\
\textbf{Exemplar-Free Class Incremental Learning: }
Exemplar-based methods~\citep{il2m,lucir,icarl,rwalk} store small subsets of data from each task. These exemplars are later replayed with current data during training for new tasks. Although effective, these methods necessitate storing input data from previous tasks, leading to multiple challenges in practical settings such as legal concerns with new regulations (e.g. European GDPR where users can request to delete personal data), and privacy issues when dealing with sensitive data like in medical signals. Recently, the
exemplar-free CIL~\citep{efcl1,efcl2,efcl3} setting has been extensively studied in the image classification domain. ADC~\citep{adc} estimates semantic drift and restores old class prototypes in the new feature space. EWC~\citep{ewc} and some more advanced versions~\citep{ewc2} calculate the importance of the parameter by the fisher information matrix then add a quadratic penalty in the loss function that penalizes the variation of each parameter to perform the previous tasks. Despite EFCIL methods are quite suitable for incremental KWS application, the exploration of EFCIL methods in KWS is limited. In addition, most EFCIL methods are only effective when starting with high-quality feature representations and always fall behind the exemplar-based methods. In this work, we propose developing a robust EFCIL method that outperforms exemplar-based approaches for small-footprint KWS applications. 
\\
\textbf{Continual learning for Speech Processing:} CL has shown promise in addressing incremental speech processing tasks by enabling systems to adapt to new data while mitigating catastrophic forgetting~\citep{cappellazzo2023sequence,yang2022online,cdoa,cl3}. Chen et al.~\citep{chen2024overcoming} proposed a hyper-gradient-based exemplar strategy for dialogue systems, periodically retraining models using selected exemplars. Xiao et al.~\citep{ucil} introduced an unsupervised framework with distillation loss to add new sound classes while maintaining task consistency. CL has also been explored for incremental KWS. RK proposed in~\citep{cl2} first introduced a diversity-based sample mechanism to select representative exemplars. More recently, DE-KWS~\citep{peng2024dark} saved model predictions to distill past knowledge beyond exemplars. However, these methods rely on storing exemplars, which creates challenges for memory- and privacy-constrained on-device applications. To this end, we propose constructing a lifelong KWS system without storing the previous predictions or data in this work. \\
\textbf{Analytic Learning.} Analytic learning (AL) uses least squares (LS) to obtain closed-form solutions, providing an efficient alternative to back propagation. Recently, the recursive formulation (e.g., BRMP~\citep{brmp}) of AL brings inspiration to CL. The BRMP can stream new samples to update the weight without weakening the impact of previous samples. ACIL~\citep{zhuang2022acil} was the first to apply AL to CL by reframing training as a recursive LS procedure, achieving accuracy similar to joint training for linear classifiers. However, our work advances AL in the speech domain by proposing the AnalyticKWS method, which adopts an exemplar-free strategy. Through recursive updates, AnalyticKWS preserves knowledge without storing any past data, representing a notable step forward for AL-based CL in speech processing. 


\begin{figure*}[t!]
\centering  
\includegraphics[width=\linewidth]{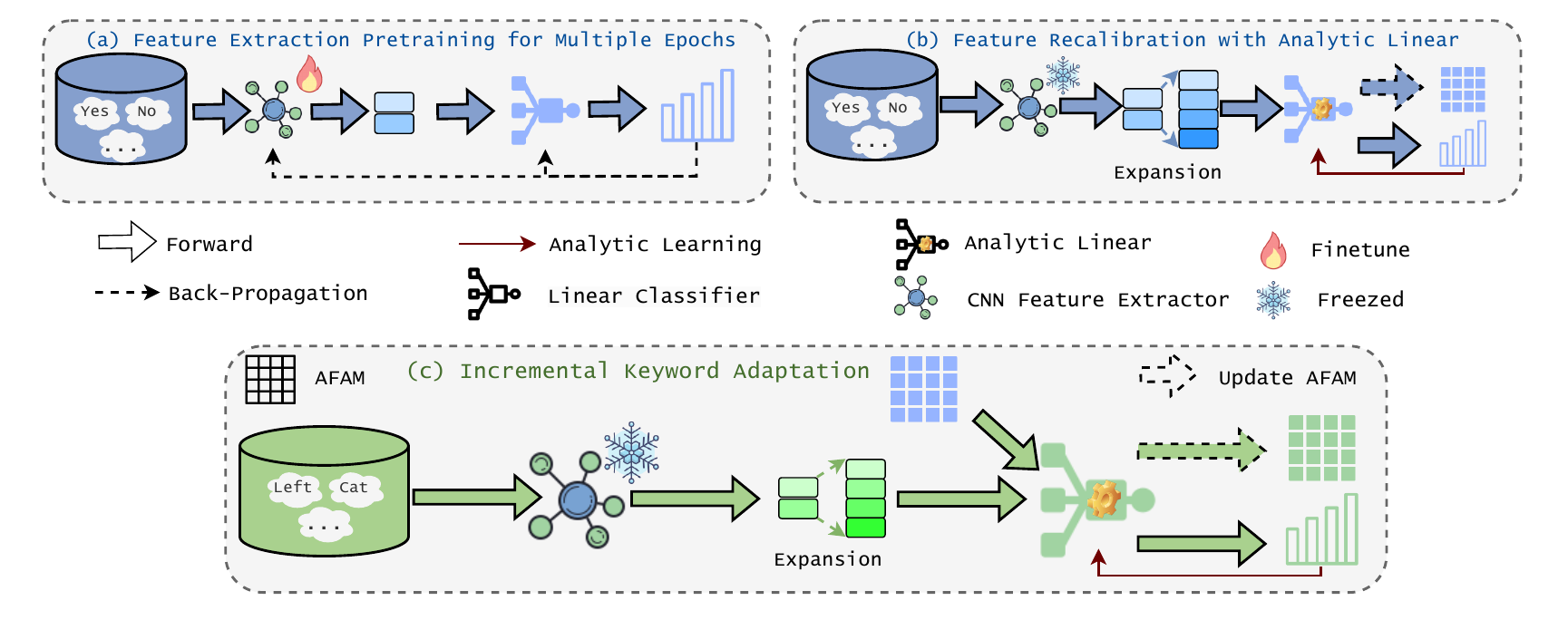}
\caption{An overview of the AnalyticKWS~\emojismiley method: (a) Train the whole model on the first task for multiple epochs to get a strong feature extractor, then (b) Apply analytic re-alignment for one epoch to increase the pre-classifier feature dimension. Next, proceed to the incremental keywords stage, where the model trains for one epoch per new task, assisted by a correlation matrix AFAM (Eq.~\eqref{eq6}) that encodes past knowledge. This process enables the model to learn new tasks while preserving previously acquired information.}
\label{fg1}
\vspace{-5mm}
\end{figure*}


\section{Our Method}
\subsection{Problem Formulation}

In this work, we examine a KWS system that learns different keyword categories through a sequence of tasks $\left \{ \tau _{0}, \tau _{1},\dots,\tau _{T} \right \}$. We treat this problem as a CIL scenario, where the system must recognize all keywords from each task, even as new tasks are introduced. For each task $\tau _{t}$, the input data $(x,y)$ follow a distinct distribution $D_{t}$. Our goal is to train a model $f(x;\theta )$ that adapts to new data while preserving its understanding of earlier tasks. Formally, we aim to minimize the cross-entropy loss across all tasks:
\begin{equation}
	\label{eq1}
	\underset{\theta}{\text{argmin}} \sum_{t = 0}^{T}\mathbb{E}_{\left ( x,y   \right )\sim D_{t}  }\left [ \mathcal{L}_{\text{CE}}\left ( y, f(x;\theta)  \right ) \right ],
\end{equation}
However, storing or reusing all past data is impractical due to memory costs and privacy concerns. Simply fine-tuning the model on new data often causes catastrophic forgetting, where the model loses the knowledge it gained from previous tasks.

\subsection{Proposed AnalyticKWS Method}
This section describes the AnalyticKWS~\emojismiley method in detail, including the feature extraction pretraining, the feature recalibration, and the incremental keyword adaptation. Our explanation focuses on small-footprint KWS models, which include a convolutional neural network (CNN) backbone as an acoustic feature extractor and a linear layer as the classifier. Figure~\ref{fg1} provides an overview of the proposed AnalyticKWS method.

\subsubsection{Feature Extraction Pretraining}
The first stage as shown in Figure~\ref{fg1}(a), is known as the feature extraction pretraining. In this step, the network is trained on the dataset $D_{0}$ of task 0 for multiple epochs using a back-propagation optimization method (e.g., stochastic gradient descent) as the conventional supervised learning to learn representations of acoustic features. After the feature extraction pretraining stage, we obtain one CNN acoustic feature extractor with weight \(\theta_{\text{cnn}}^{(0)}\) as well as one classifier with weight \(\theta_{\text{cls}}^{(0)}\). The pretrained feature extractor is then frozen to ensure consistency during subsequent stages.

\subsubsection{Feature Recalibration}

The second stage, called feature recalibration (Figure~\ref{fg1}(b)), is central to the AnalyticKWS formulation. In this step, we also use the training data $D_{0}$ (with inputs \(x_0\) and labels \(y_0\)). Unlike the last stage, we replaced the classifier with an analytic classifier to shift the network’s learning toward an analytic learning style. First, we pass the inputs through the CNN feature extractor (freezed) backbone to obtain the speech feature \(\mathbf{S}_0\).
Next, we perform an acoustic feature expansion (AFE) process by inserting an extra linear layer with weight \(\theta_{\text{afe}}\) to project \(\mathbf{S}_0\) into a higher-dimensional feature space, resulting in \(\mathbf{S}_0'\). To randomly initialize the \(\theta_{\text{afe}}\), we draw each element from a normal distribution and keep the \(\theta_{\text{afe}}\) fixed throughout training. We control the AFE by a chosen ``expansion size'' larger than the \(\mathbf{S}_0\) size. This AFE approach is very useful for small-footprint KWS because it converts the original feature into a richer representation without greatly increasing computational demands. The \(\mathbf{S}_0'\) can keep more subtle distinctions in speech signals, allowing it to preserve complex patterns. Finally, we use linear regression to map the expanded feature \(\mathbf{S}_0'\) to the label matrix \(y_0\) as:

    

\vspace{-5mm}
\begin{equation}
\label{eq2}
    \underset{{\theta}_{\text{cls}}^{(0)}}{\text{argmin}} \left|\left| {y}_0 - {\mathbf{S}}_0' {\theta}_{\text{cls}}^{(0)} \right|\right|^2_{\text{F}} + \gamma \left|\left| {\theta}_{\text{cls}}^{(0)} \right|\right|_{\text{F}}^2
\end{equation}

\noindent where \(||\cdot||_F\) indicates the Frobenius norm of matrix~\citep{fnorm}. Here we set \(\gamma\) as the regularization of Eq.~\eqref{eq2} preventing overfitting. The optimal solution to Eq.~\eqref{eq2} can be found in:
\begin{equation}
\label{eq3}
\widehat{{\theta}}_{\text{cls}}^{(0)} = \left( {\mathbf{S}}_0'{^\top} {\mathbf{S}}_0' + \gamma {I} \right)^{-1} {\mathbf{S}}_0'{^\top} {y}_0
\end{equation}

\noindent where \(\widehat{{\theta}}_{\text{cls}}^{(0)}\) indicates the estimated analytic linear layer weight of the final classifier layer before outputting the predictions. After the feature recalibration stage, the KWS model updates the classifier weights in this analytic learning style.

\subsubsection{Incremental Keyword Adaptation}

With the learning process now recalibrated to analytic learning (see Eq.~\eqref{eq3}, we can incrementally adapt to new keywords using the analytic learning approach. Suppose we can access all task data \(D_0, D_1, \dots, D_{t-1}\). In this non-continual-learning case, we can extend the learning task defined in Eq.~\eqref{eq2} to incorporate all these datasets, ensuring the model can handle multiple tasks jointly.

\begin{small}
\begin{equation}
\label{eq4}
\underset{{\theta}_{\text{cls}}^{(t-1)}}{\text{argmin}} \left|\left| 
\text{Y}_{0:t-1}
- 
\underset{0:t-1}{S'}
{\theta}_{\text{cls}}^{(t-1)}
\right|\right|_{\text{F}}^2 + \gamma \left|\left| {\theta}_{\text{cls}}^{(t-1)} \right|\right|_{\text{F}}^2
\end{equation}
\end{small}

\noindent where $\text{Y}_{0:t-1}$ is the block-diagonal matrix whose main diagonal elements are \(y_0, y_1, \dots, y_{t-1}\). And \(\underset{0:t-1}{S'}\) is formed by stacking the expanded feature matrices. The solution to Eq.\eqref{eq4} can be written as:



\begin{small}
\begin{equation}
\widehat{{\theta}}_{\text{cls}}^{(t-1)}  = 
 \left( 
\sum_{i=0}^{t-1} {\mathbf{S}}_i'{^\top} {\mathbf{S}}_i' + \gamma {I} \right)^{-1} \underset{0:t-1}{{\mathbf{S}}'{^\top}} \text{Y}_{0:t-1}
\label{eq7}
\end{equation}
\end{small}

\noindent where \(\widehat{{\theta}}_{\text{cls}}^{(t-1)}\) with a column size proportional to task number \(t\). The goal of AnalyticKWS is
to calculate the analytical solution that satisfies~\eqref{eq4} at task \(\tau_{t}\) based on \(\widehat{{\theta}}_{\text{cls}}^{(t-1)}\) given \(D_{t}\). Specifically, we aim to obtain \(\widehat{{\theta}}_{\text{cls}}^{(t)}\) recursively based on \(\widehat{{\theta}}_{\text{cls}}^{(t-1)}\), \(\mathbf{S}_{t}'\), and label \(y_t\) that are available only at the current task. When the updated weight \(\widehat{{\theta}}_{\text{cls}}^{(t)}\) satisfy Eq.~\eqref{eq4} with all previous task data, AnalyticKWS could reduce forgetting in the sense that the recursive formulation (i.e., incremental learning) gives the same answer with the joint learning. To achieve this, we introduce \(\mathbb{A}_{t-1}\), the acoustic feature autocorrelation matrix (AFAM) from the task \(\tau_{t-1}\).


\vspace{-2mm}
\begin{equation}
\label{eq6}
\mathbb{A}_{t-1} = \left( \sum_{i=0}^{t-1} \mathbf{S}_i'{^\top} \mathbf{S}_i' + \gamma I \right)^{-1}
\end{equation}


\noindent With the  the weight \(\widehat \theta_{\text{cls}}^{(t)}\) could obtained by: 

\vspace{-4mm}
            \begin{small}
		\begin{align}\label{eq_w_update}
			 \widehat \theta_{\text{cls}}^{(t)}
			=\begin{bmatrix} \widehat \theta_{\text{cls}}^{(t-1)} -  \mathbb{A}_{t}{\mathbf{S}}_{t}^{\top}\mathbf{S}_{t}' \widehat \theta_{\text{cls}}^{(t-1)}\ \ \   \mathbb{A}_{t}\mathbf{S}_t'{^\top}y_t \end{bmatrix}
		\end{align}
            \end{small}
            
\noindent which is identical to that obtained by \eqref{eq7}.
To calculated the weight, the current AFAM $\mathbb{A}_{t}$ can also be recursively calculated by:

\vspace{-4mm}
\begin{equation}
\label{delta}
	\Delta = \mathbb{A}_{t-1} \mathbf{S}_t'{^\top} (I + \mathbf{S}_t'\mathbb{A}_{t-1} \mathbf{S}_t'{^\top})^{-1} \mathbf{S}_t'{^\top} \mathbb{A}_{t-1}
\end{equation}    
\vspace{-4mm}
\begin{equation}
\label{eq_R_update}
\mathbb{A}_{t} = \mathbb{A}_{t-1} - \Delta
\end{equation}

\noindent For the full proof please see the appendix.



As a result, the final classifier layer weight can be updated recursively using \(\widehat{\theta}_{\text{cls}}^{(t-1)}\), \(\mathbf{S}_{t}'\), \(\mathbb{A}_{t}\) and label \(y_t\). This means that even though the KWS model is incremental learning of incoming keywords, the classifier prediction is equal to the outcome of a joint analytic learning solution applied to all tasks. 



\begin{algorithm}[t]
	\caption{AnalyticKWS~\emojismiley}
	\label{alg:acil}
		\begin{algorithmic}
			\STATE {\bfseries Feature Extraction Pretraining:} with $D_{0}$. Conventional supervised training for multiple epochs on the task 0.\\
            			\quad\\
			\STATE {\bfseries Feature Recalbration:} \STATE \textcolor{blue}{i)} Obtain expanded feature matrix with AFE; \STATE \textcolor{blue}{ii)} Obtain re-aligned weight $\widehat \theta_{\text{cls}}^{(0)}$ with \eqref{eq3}.  \STATE \textcolor{blue}{iii)} Save feature autocorrelation matrix $\mathbb{A}_{0}$.\\
			\quad\\
			\STATE {\bfseries Incremental Keyword Adaptation:}
			\FOR{$t=1$ {\bfseries to} $T$ (with $D_{t}$, $\widehat \theta_{\text{cls}}^{(t-1)}$ and  $\mathbb{A}_{t-1}$)}
			\STATE \textcolor{blue}{i)} Obtain and stack the feature matrix;
			\STATE \textcolor{blue}{ii)} Update $\mathbb{A}_{t}$ with \eqref{delta} and \eqref{eq_R_update};
			\STATE \textcolor{blue}{iii)} Update weight matrix $\widehat \theta_{\text{cls}}^{(t)}$ with \eqref{eq_w_update};
			\ENDFOR
		\end{algorithmic}
\end{algorithm}

We summarize the computational steps of AnalyticKWS in Alg.~\ref{alg:acil}. This algorithm begins with a Feature Extraction Pretraining, where the model first learns from the dataset using conventional back-propagation training. After this training, we freeze the feature extractor. Then we input the data of task 0 again for the Feature Recalibration. 
We first utilize the AFE to obtain the speech feature matrix. Then based on the feature matrix, we shift the classifier into analytic learning and save the acoustic feature autocorrelation matrix. Following the recalibration stage, the algorithm moves into class incremental learning. AnalyticKWS uses the newly received utterances for the new keywords in each task, extracts its feature matrix, updates the AFAM, and finally updates the linear classifier weight. This process is repeated for each incoming task, ensuring the model adapts to new keywords while preserving knowledge from all previously learned tasks.



\begin{table*}[t]
\centering
\caption{Comparison of various CL methods for KWS. Finetune serves as the lower bound, and Joint training acts as the upper bound. We evaluate each method on accuracy (ACC in \%) and backward transfer (BWT). ``T" is the task number. \textbf{Bold} values indicate the best results, and \uline{underlined} values denote the second-best. A dash (-) marks unavailable results. An asterisk (*) signifies that the method uses a buffer of size 500 for exemplar storage. The proposed AnalyticKWS methods are \colorbox[HTML]{C4D5EB}{highlighted}.}
\label{tab:my-table}
\resizebox{0.9\textwidth}{!}{%
\begin{tabular}{c|l|ccc|ccc|ccc}
\toprule
\multirow{2}{*}{\textbf{Metric}} &
  \multirow{2}{*}{\textbf{Method}} &
  \multicolumn{3}{c|}{\textbf{GSC-v1}} &
  \multicolumn{3}{c|}{\textbf{GSC-v2}} &
  \multicolumn{3}{c}{\textbf{SC-100}} \\
 &
   &
  \textbf{T=6} &
  \textbf{T=11} &
  \textbf{T=21} &
  \textbf{T=6} &
  \textbf{T=11} &
  \textbf{T=21} &
  \textbf{T=11} &
  \textbf{T=26} &
  \textbf{T=51} \\ \midrule
\multirow{10}{*}{\textbf{ACC(\%$\uparrow$)}} & Joint   &  \multicolumn{3}{c|}{94.93} &
  \multicolumn{3}{c|}{94.76} &
  \multicolumn{3}{c}{95.32} \\

& Finetune & 26.84  & 17.99  & 9.59   & 30.07  & 16.82  & 8.92   & 15.07  & 6.45   & 3.30    \\
                               
                               & EWC~\citep{ewc}      & 72.28  & 71.65  & 69.66  & 71.55  & 68.20   & 66.76  & 43.90   & 40.56  & 35.39  \\
                               & BiC*~\citep{bic}      & 80.22  & 79.39  & 79.19  & 75.79  & 76.52  & 76.92  & \multicolumn{3}{c}{-}    \\
                               & iCaRL*~\citep{icarl}    & 85.24  & 81.14  & 73.61  & 84.72  & 79.16  & 67.35  & 69.3   & 46.34  & 23.70   \\
                               & Rwalk*~\citep{rwalk}    & 87.03  & 85.38  & 84.55  & 87.12  & 87.27  & 86.77  & 76.93  & 77.21  & 76.78  \\
                               & RK*~\citep{cl2}       & 85.56  & 83.19  & 80.87  & 83.49  & 80.52  & 78.91  & 68.72  & 61.62  & 59.54  \\
                               & DE-KWS*~\citep{peng2024dark}   & 88.82  & \uline{85.59}  & \uline{85.53}  & 87.78  & 85.34  & 82.38  & 67.71  & 59.78  & 54.34  \\
                               & \cellcolor[HTML]{C4D5EB}AnalyticKWS-128 & \cellcolor[HTML]{C4D5EB}\uline{88.95}  & \cellcolor[HTML]{C4D5EB}84.91  & \cellcolor[HTML]{C4D5EB}84.58  & \cellcolor[HTML]{C4D5EB}\uline{88.88}  & \cellcolor[HTML]{C4D5EB}\uline{88.87}  & \cellcolor[HTML]{C4D5EB}\uline{88.85}  & \cellcolor[HTML]{C4D5EB}\uline{85.77}  & \cellcolor[HTML]{C4D5EB}\uline{85.66}  & \cellcolor[HTML]{C4D5EB}\uline{85.55}  \\
                               & \cellcolor[HTML]{C4D5EB}AnalyticKWS-256 & \cellcolor[HTML]{C4D5EB}\textbf{89.51}  & \cellcolor[HTML]{C4D5EB}\textbf{85.83}   & \cellcolor[HTML]{C4D5EB}\textbf{85.60}   & \cellcolor[HTML]{C4D5EB}\textbf{89.48}  & \cellcolor[HTML]{C4D5EB}\textbf{89.53}  & \cellcolor[HTML]{C4D5EB}\textbf{89.50}   & \cellcolor[HTML]{C4D5EB}\textbf{87.99}  & \cellcolor[HTML]{C4D5EB}\textbf{87.85}  & \cellcolor[HTML]{C4D5EB}\textbf{87.63}  \\ \midrule
\multicolumn{1}{l|}{\multirow{10}{*}{\textbf{BWT$(\uparrow)$}}} & Joint   &  \multicolumn{3}{c|}{-} &
  \multicolumn{3}{c|}{-} &
  \multicolumn{3}{c}{-} \\ &
  Finetune &
  -0.376 &
  -0.249 &
  -0.163 &
  -0.362 &
  -0.256 &
  -0.166 &
  -0.264 &
  -0.144 &
  -0.086 \\
\multicolumn{1}{l|}{}          & EWC~\citep{ewc}      & -0.117 & -0.061 & -0.035 & -0.122 & -0.072 & -0.045 & -0.146 & -0.076 & -0.048 \\
\multicolumn{1}{l|}{}          & BiC*~\citep{bic}      & -0.084 & -0.045 & -0.025 & -0.095 & -0.053 & -0.028 & \multicolumn{3}{c}{-}    \\
\multicolumn{1}{l|}{}          & iCaRL*~\citep{icarl}    & -0.054 & -0.037 & -0.029 & -0.057 & -0.041 & -0.032 & -0.067 & -0.047 & -0.038 \\
\multicolumn{1}{l|}{}          & Rwalk*~\citep{rwalk}    & -0.048 & -0.026 & -0.015 & -0.047 & -0.024 & -0.014 & -0.052 & -0.023 & -0.013 \\
\multicolumn{1}{l|}{}          & RK*~\citep{cl2}       & -0.047 & -0.033 & -0.021 & -0.061 & -0.040  & -0.025 & -0.065 & -0.040  & -0.023 \\
\multicolumn{1}{l|}{}          & DE-KWS*~\citep{peng2024dark}   & \textbf{-0.032} & -0.026  & -0.014 & -0.037 & -0.024 & -0.015 & -0.058 & -0.030  & -0.018 \\
\multicolumn{1}{l|}{}          & \cellcolor[HTML]{C4D5EB}AnalyticKWS-128 & \cellcolor[HTML]{C4D5EB}\uline{-0.034} & \cellcolor[HTML]{C4D5EB}\uline{-0.025} & \cellcolor[HTML]{C4D5EB}\uline{-0.013} & \cellcolor[HTML]{C4D5EB}\uline{-0.033} & \cellcolor[HTML]{C4D5EB}\uline{-0.016} & \cellcolor[HTML]{C4D5EB}\uline{-0.008} & \cellcolor[HTML]{C4D5EB}\uline{-0.021} & \cellcolor[HTML]{C4D5EB}\uline{-0.008} & \cellcolor[HTML]{C4D5EB}\uline{-0.004} \\
\multicolumn{1}{l|}{}          & \cellcolor[HTML]{C4D5EB}AnalyticKWS-256 & \cellcolor[HTML]{C4D5EB}\textbf{-0.032} & \cellcolor[HTML]{C4D5EB}\textbf{-0.024} & \cellcolor[HTML]{C4D5EB}\textbf{-0.012} & \cellcolor[HTML]{C4D5EB}\textbf{-0.030}  & \cellcolor[HTML]{C4D5EB}\textbf{-0.015} & \cellcolor[HTML]{C4D5EB}\textbf{-0.007} & \cellcolor[HTML]{C4D5EB}\textbf{-0.017} & \cellcolor[HTML]{C4D5EB}\textbf{-0.007} & \cellcolor[HTML]{C4D5EB}\textbf{-0.003} \\ \bottomrule
\end{tabular}%
}
\end{table*}

\section{Experiment Setting}
\subsection{Dataset}
Unlike previous CL studies on KWS that focus on a single dataset, we evaluate our method on three different datasets to show its robustness. First, we use the widely adopted Google Speech Commands (GSC) v1 dataset, which includes 64,727 short audio clips, each lasting one second, covering 30 distinct keywords. We also use the larger GSC v2 dataset with 105,829 audio clips. This expanded version contains the original 30 keywords plus 5 new words (``Backward", ``Forward", ``Follow", ``Learn", and   ``Visual"), resulting in a richer variety of speakers and improved data diversity. Following established practices, we split each dataset into training (80\%) and validation (20\%) sets, with all audio sampled at 16 kHz.

In addition, we evaluate our method on the SC-100 dataset~\citep{sc100}, which consists of 313,951 keyword utterances covering 100 different keywords. The SC-100 dataset is created from the LibriSpeech corpus using the KeywordMiner tool, which identifies words and their timestamps, and a segmenter that extracts individual words from full sentences. This process results in a large, diverse dataset suitable for complex KWS tasks.

Following~\citep{peng2024dark,zhuang2022acil}, we first train the network (Task 0) on a base dataset. Then, the network learns the remaining classes over \(T\) tasks, with each phase containing classes disjoint from earlier tasks. For the GSC dataset, we report results for \(T = 6,11,21\). As an example, when \(T=11\), we pre-train TC-ResNet-8 using 10 unique keywords from GSC-v1; the remaining data is divided into 20 tasks, each holding 1 new keyword. For SC-100, we extend \(T\) to 51, with 50 keywords for the base training phase and 50 follow-up tasks to test large-scale incremental learning. For more details please see the appendix.

\subsection{Experimental Setup}
We use 40-dimensional MFCC with a 160 hop length as input features and adopt the TC-ResNet-8 model as the backbone following~\citep{peng2024dark}. TC-ResNet-8~\citep{choi2019temporal} is a lightweight CNN developed for KWS on devices with limited computing. It contains three residual blocks, each composed of 1D temporal convolutional layers, batch normalization layers, and ReLU activation functions. Across these layers, the channel sizes are $ \left \{16,24,32,48 \right \}$, including the first convolutional layer. For each task, we train the model for 50 epochs. The suffix "X" in AnalyticKWS-X refers to the dimensionality of the feature space after applying the AFE. 

\subsection{Metrics}
We first use two metrics for performance evaluation: Average Accuracy (ACC), and Backward Transfer (BWT)~\citep{lopez2017gradient}. ACC is the average accuracy over all completed tasks that evaluates the overall performance of CIL algorithms: $\text{ACC} = \frac{1}{T+1}{\sum}_{t=0}^{T}\text{A}_{t}$ where $\text{A}_{t}$ indicates the average test accuracy of the network incrementally trained at task $t$ by testing it on $D_{0:t}^{\text{test}}$. A higher $\text{ACC}$ score is preferred when evaluating CL algorithms. BWT measures how learning new tasks affects previous tasks: $\text{BWT} = \frac{1}{T} \sum_{t=1}^{T} \left( \text{A}_{T} -\text{A}_{t} \right)$ where $\text{A}_T$ represents the final average accuracy after all $T$ tasks are learned. A positive $\text{BWT}$ suggests that learning new tasks improves performance on earlier tasks, while a negative $\text{BWT}$ indicates catastrophic forgetting. We also assess efficiency using task training time (TT) and extra memory. TT is the average time required to train each epoch of all tasks. Extra memory represents the extra memory used to store replay data or model weights.
\begin{figure*}[t]
  \centering
  \includegraphics[width=0.9\linewidth]{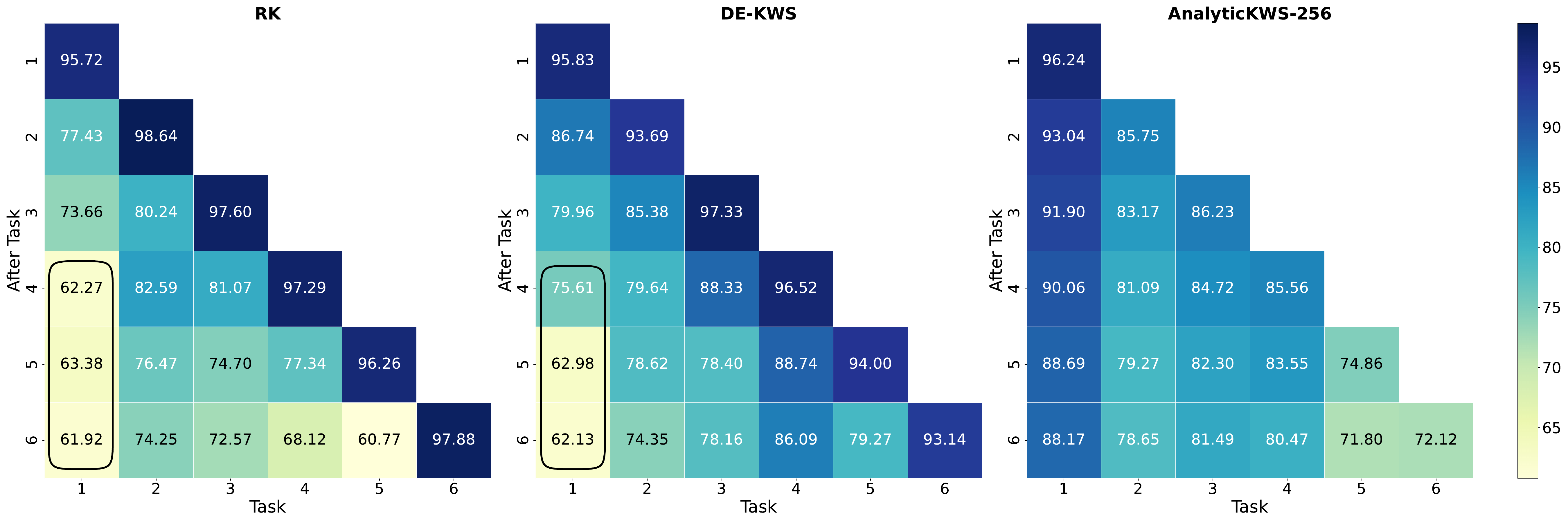 }
  \vspace{-1mm}
  \caption{Task-wise performance comparison of different methods with 500 buffer size.}
  \label{fig:heatmap}
  \vspace{-3mm}
\end{figure*}

\section{Results and Analysis}

\subsection{Comparable Study for ACC\&BWT}
Table~\ref{tab:my-table} compares various CL methods for KWS, using ACC and BWT as key metrics. In these experiments, Finetune represents the lower bound, while Joint training serves as the upper bound. The Finetune method suffers from significant forgetting and achieves low ACC with high negative BWT. EWC and BiC show some improvement, but not very significant. Other exemplar-based methods, such as iCaRL, Rwalk, and RK, maintain better ACC due to storing examples in a buffer (size = 500), but this practice adds extra memory usage. DE-KWS is also an exemplar-based baseline that achieves reasonable accuracy as the most recent baseline, yet it still does not match the best results. In contrast, AnalyticKWS-128 and AnalyticKWS-256 achieve stronger and more consistent ACC across the tasks and datasets. They exhibit minimal forgetting, as shown by their higher BWT scores, often approaching the ideal performance of Joint training. Crucially, these methods do not use exemplars, preserving data privacy and cutting down on memory needs. Overall, the results demonstrate the effectiveness of our AnalyticKWS method for continual KWS. It offers near-Joint accuracy without needing a large exemplar buffer, proving that our approach can mitigate catastrophic forgetting and maintain high performance.

\subsection{Comparable Study for TT}
\begin{table}[t!]
\centering
\caption{Average task training times TT (Second) comparison across methods. Each method is evaluated based on the average (Avg.) TT across three settings.}
\resizebox{0.9\linewidth}{!}{%
\begin{tabular}{l|c|c|c}
\toprule
\textbf{Method}   & \textbf{GSC-v1 Avg.} & \textbf{GSC-v2 Avg.} & \textbf{SC-100 Avg.} \\ \midrule
Finetune          & 262.08               & 277.75               & 433.29                           \\
EWC              & 373.54               & 454.10               & 827.21                              \\
BiC             & 288.67               & 372.51               & -                                   \\
iCaRL           & 353.04               & 410.81               & 453.33                              \\
Rwalk             & 385.13               & 538.16               & 865.59                              \\
RK               & 956.55               & 1239.46              & 1771.76                             \\
DE-KWS           & 270.82               & 350.42               & 576.85                             \\
\cellcolor[HTML]{C4D5EB}AnalyticKWS-128 & \cellcolor[HTML]{C4D5EB}\textbf{5.09}        & \cellcolor[HTML]{C4D5EB}\textbf{5.97}        & \cellcolor[HTML]{C4D5EB}\textbf{9.31}                \\
\cellcolor[HTML]{C4D5EB}AnalyticKWS-256 & \cellcolor[HTML]{C4D5EB}\uline{5.49}        & \cellcolor[HTML]{C4D5EB}\uline{6.48}        & \cellcolor[HTML]{C4D5EB}\uline{10.47}               \\ \bottomrule
\end{tabular}}
\label{tab:time_efficiency}
\vspace{-5mm}
\end{table}

Table~\ref{tab:time_efficiency} shows that our proposed AnalyticKWS reduces TT across all datasets, allowing faster learning of new tasks. We calculate the training time per epoch as the TT. All experiments are estimated by the NVIDIA RTX 3090. Methods like EWC, Rwalk, and RK demand more computation because they track extra parameters or buffers. DE-KWS also has a lower TT than some baselines but still cannot match AnalyticKWS. In contrast, AnalyticKWS-128 and AnalyticKWS-256 reach higher efficiency without storing large numbers of examples and only require one epoch to adapt to each new task. As a result, they operate more efficiently, running faster and consuming fewer resources on small-footprint devices.

\subsection{Comparable Study for Extra Memory}
The AnalyticKWS stores \(\mathbb{A}_{t}\) instead of speech clips or the previous model weights. As an example, the memory used by storing AnalyticKWS-128 on all three datasets is \(128\times128=16\)K tensor elements, while other methods consume 8M (e.g.,on GSC-v1 with 500 buffer is at least \(16000\times1\times1\times500\approx8\)M). Some methods like Rwalk and RK even require preserving the whole weight of the existing model. With a limited buffer size but large task numbers, the rehearsal-based method performs struggles in SC-100. This shows that our method is memory-friendly to large-scale KWS datasets (e.g., SC-100) in the edge-device application for example the robot voice control.
\begin{figure}[t!]
    \centering
    \vspace{-10mm}
    \includegraphics[width=\linewidth]{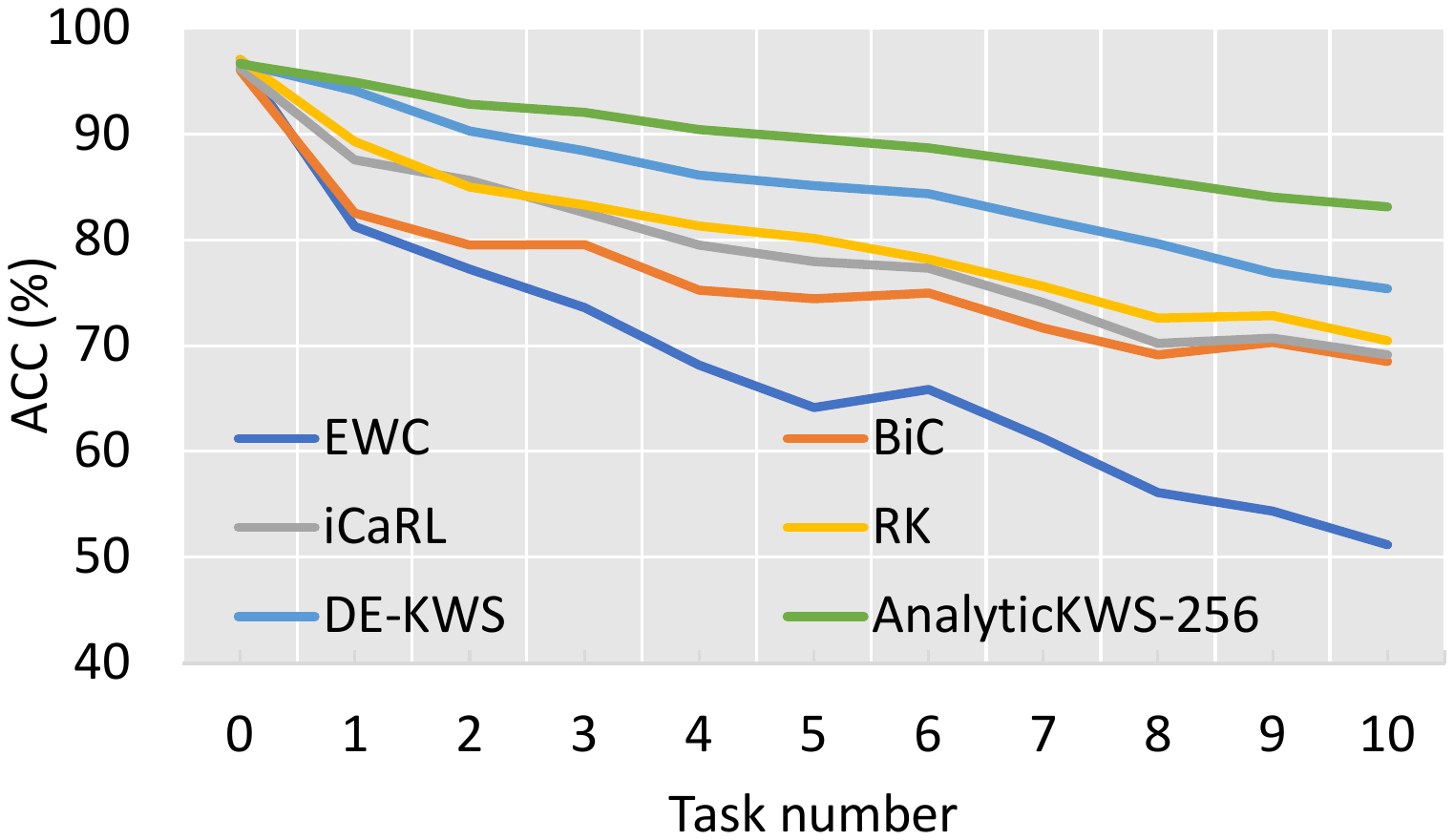}
    \vspace{-14mm}
    \caption{Task-wise accuracy on GSC-v2 with 11 tasks.}
    \label{fig:enter-label1}
    \vspace{-5mm}
\end{figure}

\begin{figure}[t!]
    \centering
    \vspace{-5mm}
    \includegraphics[width=\linewidth]{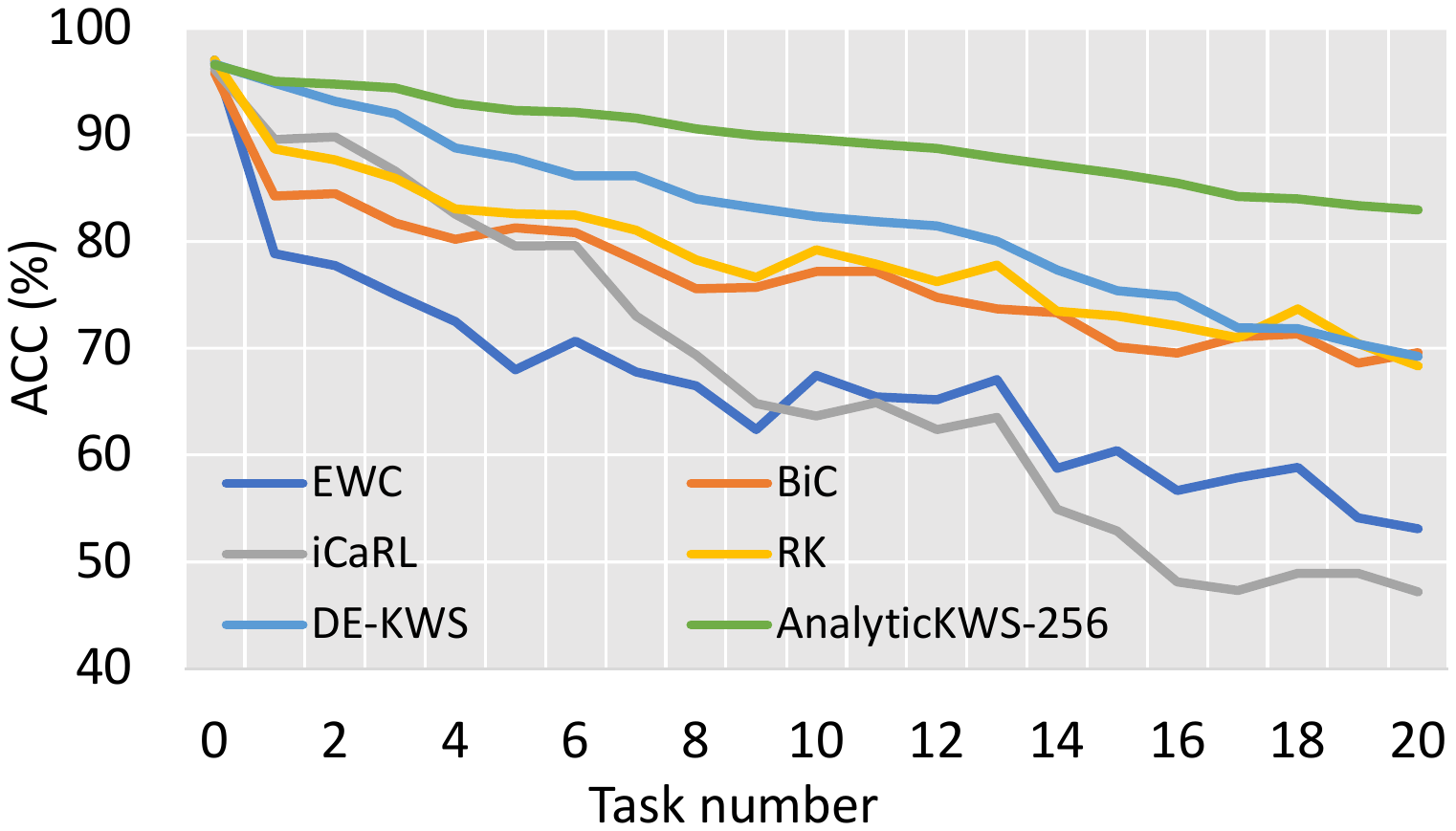}
    \vspace{-14mm}
    \caption{Task-wise accuracy on GSC-v2 with 21 tasks.}
    \label{fig:enter-label2}
    \vspace{-5mm}
\end{figure}

\section{Task-wise Analysis}
From the heatmap in Figure~\ref{fig:heatmap} (GSC-v2, six tasks), we observe that DE-KWS maintains high accuracy in early and later tasks through its “dark experience” strategy, effectively balancing long-term retention and new-task adaptation. However, our AnalyticKWS shows better stability and accuracy across the entire task sequence, despite using no buffer for replay. In contrast, approaches like RK—which stores 500 exemplars—still struggle with mid-term forgetting (e.g., Task 2), suggesting that their reliance on extra data does not guarantee sustained performance. AnalyticKWS avoids storing historical samples but remains resistant to catastrophic forgetting through its analytic learning updates, enabling it to preserve key information from past tasks while smoothly integrating new ones.

As illustrated in Figures~\ref{fig:enter-label1} and~\ref{fig:enter-label2}, task-wise accuracy on GSC-v2 steadily declines as the task count grows from 1 to 11 and then up to 21, highlighting the difficulty of preventing catastrophic forgetting over many tasks. Methods such as EWC, BiC, and RK drop quickly as they learn more classes, indicating a struggle to maintain old knowledge. Notably, iCaRL faces only a moderate drop at 11 tasks but suffers a much steeper decline at 21 tasks, likely because its fixed-size buffer cannot store enough representative exemplars for a larger number of classes, leading to greater forgetting. While DE-KWS performs better than these baselines, it still undergoes a downward trend across tasks. By contrast, AnalyticKWS-256 preserves higher accuracy in both 11-task and 21-task scenarios, suggesting that its exemplar-free, analytic approach more effectively balances long-term retention and new-class adaptation.

\section{Ablation Study}
This ablation study compares models with different acoustic feature expansion (AFE) sizes and regularization settings as reported in Table~\ref{tab:abl}. Without AFE and only regularization, the accuracy is 86.57\%. As we introduce a small AFE (64) with regularization, the accuracy improves to 87.19\%, and further expansion from 128 to 512 dimensions continues to enhance performance, reaching a peak accuracy of 89.68\% with the 512-dimensional AFE. Removing regularization at this level slightly decreases accuracy to 89.64\%. These findings confirm that combining an expanded feature space with regularization is crucial to maximize accuracy, while models lacking either approach exhibit lower performance. This result demonstrates the effectiveness of our proposed method.


\begin{table}[t!]
\centering
\caption{Ablation study of acoustic feature expansion (AFE) and regularization in AnalyticKWS. The symbol ``\cmark'' indicates the use of AFE or regularization, while ``\xmark'' means they are disabled. Accuracy (ACC) improves by increasing the AFE size and combining it with regularization, with the best result obtained by a 512-dimensional expansion plus regularization.}
\vspace{-2mm}
\resizebox{0.9\columnwidth}{!}{%
\begin{tabular}{c|c|c}
\toprule
\textbf{Feature Expansion} & \textbf{Regularization} & \textbf{ACC$(\% \uparrow)$}     \\ \midrule
\xmark                 & \cmark             & 86.57 \\
\cmark (64)            & \cmark             & 87.19 \\
\cmark (128)           & \cmark             & 88.72 \\
\cmark (256)           & \cmark             & 89.23 \\
\cmark (512)           & \cmark             & \textbf{89.68} \\
\cmark (512)           & \xmark              & 89.64 \\ \bottomrule
\end{tabular}%
}
\label{tab:abl}
\vspace{-5mm}
\end{table}

\section{Conclusion}
In this work, we have introduced a novel exemplar-free analytic CL method, namely AnalyticKWS that addresses catastrophic forgetting and protects data privacy by avoiding the storage of historical examples. Incorporating a closed-form analytic update, our approach maintains knowledge across multiple tasks and ensures that incremental learning matches the performance of joint training without requiring repeated access to old data. The recursive structure of AnalyticKWS grants absolute memorization, allowing it to achieve state-of-the-art results in both small-scale and large-phase scenarios. Our experiments on various KWS benchmarks verify these benefits, highlighting AnalyticKWS’s potential for practical deployment on resource-limited devices.

\section*{Limitations}
While our proposed AnalyticKWS method is privacy-preserving and shows strong performance, it still has some limitations. First, we have not explored its effectiveness in multilingual KWS, which remains a vital challenge for real-world speech applications. Second, the current CNN-based feature extractor, used similarly to transfer learning, might not be optimal for every domain, and improving it could increase the computational costs of GPU operations. Lastly, although AnalyticKWS retains knowledge well, enhancing its plasticity for future learning is necessary for scenarios that demand rapid task switching or adaptation.
\section*{Ethics Statement}
All the data used in this paper are publicly available and are used under the following licenses: the Creative Commons BY-NC-ND 4.0 License and Creative Commons Attribution 4.0 International License, the TED Terms of Use, the YouTube’s Terms of Service, and the BBC’s Terms of Use. 
\bibliography{main}

\begin{thebibliography}{43}
\providecommand{\natexlab}[1]{#1}

\bibitem[{Awasthi et~al.(2021)Awasthi, Kilgour, and Rom}]{awasthi2021teaching}
Abhijeet Awasthi, Kevin Kilgour, and Hassan Rom. 2021.
\newblock Teaching keyword spotters to spot new keywords with limited examples.
\newblock In \emph{Proc. Interspeech}.

\bibitem[{Bello et~al.(2018)Bello, Mydlarz, and Salamon}]{smarthome}
Juan~Pablo Bello, Charlie Mydlarz, and Justin Salamon. 2018.
\newblock {Sound Analysis in Smart Cities}.
\newblock \emph{Springer International Publishing}, pages 373--397.

\bibitem[{Belouadah and Popescu(2019)}]{il2m}
Eden Belouadah and Adrian Popescu. 2019.
\newblock Il2m: Class incremental learning with dual memory.
\newblock In \emph{Proc. the IEEE/CVF International Conference on Computer Vision}, pages 583--592.

\bibitem[{Belouadah et~al.(2021)Belouadah, Popescu, and Kanellos}]{cil1}
Eden Belouadah, Adrian Popescu, and Ioannis Kanellos. 2021.
\newblock {A Comprehensive Study of Class Incremental Learning Algorithms for Visual Tasks}.
\newblock \emph{Neural Networks}, 135:38--54.

\bibitem[{Cappellazzo et~al.(2023)Cappellazzo, Yang, Falavigna, and Brutti}]{cappellazzo2023sequence}
Umberto Cappellazzo, Muqiao Yang, Daniele Falavigna, and Alessio Brutti. 2023.
\newblock Sequence-level knowledge distillation for class-incremental end-to-end spoken language understanding.
\newblock \emph{Proceedings of Interspeech}.

\bibitem[{Chaudhry et~al.(2018)Chaudhry, Dokania, Ajanthan, and Torr}]{rwalk}
Arslan Chaudhry, Puneet~K Dokania, Thalaiyasingam Ajanthan, and Philip~HS Torr. 2018.
\newblock Riemannian walk for incremental learning: Understanding forgetting and intransigence.
\newblock In \emph{Proceedings of the European conference on computer vision (ECCV)}, pages 532--547.

\bibitem[{Chen et~al.(2024)Chen, Li, Hu, Chen, Qin, and Zhang}]{chen2024overcoming}
Chen Chen, Ruizhe Li, Yuchen Hu, Yuanyuan Chen, Chengwei Qin, and Qiang Zhang. 2024.
\newblock Overcoming catastrophic forgetting by exemplar selection in task-oriented dialogue system.
\newblock \emph{arXiv preprint arXiv:2405.10992}.

\bibitem[{Chen et~al.(2014)Chen, Parada, and Heigold}]{chen2014small}
Guoguo Chen, Carolina Parada, and Georg Heigold. 2014.
\newblock Small-footprint keyword spotting using deep neural networks.
\newblock In \emph{Proc. IEEE International Conference on Acoustics, Speech and Signal Processing (ICASSP)}, pages 4087--4091.

\bibitem[{Choi et~al.(2019)Choi, Seo, Shin, Byun, Kersner, Kim, Kim, and Ha}]{choi2019temporal}
Seungwoo Choi, Seokjun Seo, Beomjun Shin, Hyeongmin Byun, Martin Kersner, Beomsu Kim, Dongyoung Kim, and Sungjoo Ha. 2019.
\newblock Temporal convolution for real-time keyword spotting on mobile devices.
\newblock In \emph{Proc. Interspeech}.

\bibitem[{Golub and Van~Loan(2013)}]{fnorm}
Gene~H Golub and Charles~F Van~Loan. 2013.
\newblock \emph{Matrix computations}.
\newblock JHU press.

\bibitem[{Goswami et~al.(2024{\natexlab{a}})Goswami, Liu, Twardowski, and van~de Weijer}]{szatkowski2024adapt}
Dipam Goswami, Yuyang Liu, Bart Twardowski, and Joost van~de Weijer. 2024{\natexlab{a}}.
\newblock Fecam: Exploiting the heterogeneity of class distributions in exemplar-free continual learning.
\newblock \emph{Advances in Neural Information Processing Systems}, 36.

\bibitem[{Goswami et~al.(2024{\natexlab{b}})Goswami, Liu, Twardowski, and van~de Weijer}]{efcl3}
Dipam Goswami, Yuyang Liu, Bart{\l}omiej Twardowski, and Joost van~de Weijer. 2024{\natexlab{b}}.
\newblock Fecam: Exploiting the heterogeneity of class distributions in exemplar-free continual learning.
\newblock \emph{Advances in Neural Information Processing Systems}, 36.

\bibitem[{Goswami et~al.(2024{\natexlab{c}})Goswami, Soutif-Cormerais, Liu, Kamath, Twardowski, van~de Weijer et~al.}]{goswami2024resurrecting}
Dipam Goswami, Albin Soutif-Cormerais, Yuyang Liu, Sandesh Kamath, Bart Twardowski, Joost van~de Weijer, et~al. 2024{\natexlab{c}}.
\newblock Resurrecting old classes with new data for exemplar-free continual learning.
\newblock In \emph{Proceedings of the IEEE/CVF Conference on Computer Vision and Pattern Recognition}, pages 28525--28534.

\bibitem[{Goswami et~al.(2024{\natexlab{d}})Goswami, Soutif-Cormerais, Liu, Kamath, Twardowski, van~de Weijer et~al.}]{adc}
Dipam Goswami, Albin Soutif-Cormerais, Yuyang Liu, Sandesh Kamath, Bart Twardowski, Joost van~de Weijer, et~al. 2024{\natexlab{d}}.
\newblock Resurrecting old classes with new data for exemplar-free continual learning.
\newblock In \emph{Proceedings of the IEEE/CVF Conference on Computer Vision and Pattern Recognition}, pages 28525--28534.

\bibitem[{Hou et~al.(2019)Hou, Pan, Loy, Wang, and Lin}]{lucir}
Saihui Hou, Xinyu Pan, Chen~Change Loy, Zilei Wang, and Dahua Lin. 2019.
\newblock {Learning a Unified Classifier Incrementally via Rebalancing}.
\newblock In \emph{Proc. IEEE/CVF Conference on Computer Vision and Pattern Recognition (CVPR)}, pages 831--839.

\bibitem[{Huang et~al.(2022)Huang, Hou, and Chen}]{pclkws2022}
Yizheng Huang, Nana Hou, and Nancy~F Chen. 2022.
\newblock Progressive continual learning for spoken keyword spotting.
\newblock In \emph{Proc. IEEE International Conference on Acoustics, Speech and Signal Processing (ICASSP)}, pages 7552--7556.

\bibitem[{Kim et~al.(2021)Kim, Chang, Lee, and Sung}]{kim2021broadcasted}
Byeonggeun Kim, Simyung Chang, Jinkyu Lee, and Dooyong Sung. 2021.
\newblock Broadcasted residual learning for efficient keyword spotting.
\newblock In \emph{Proc. Interspeech}.

\bibitem[{Kirkpatrick et~al.(2017)Kirkpatrick, Pascanu, Rabinowitz, Veness, Desjardins, Rusu, Milan, Quan, Ramalho, Grabska-Barwinska et~al.}]{ewc}
James Kirkpatrick, Razvan Pascanu, Neil Rabinowitz, Joel Veness, Guillaume Desjardins, Andrei~A Rusu, Kieran Milan, John Quan, Tiago Ramalho, Agnieszka Grabska-Barwinska, et~al. 2017.
\newblock {Overcoming Catastrophic Forgetting in Neural Networks}.
\newblock \emph{Proceedings of the National Academy of Sciences}, 114(13):3521--3526.

\bibitem[{L{\'o}pez-Espejo et~al.(2021)L{\'o}pez-Espejo, Tan, Hansen, and Jensen}]{lopezespejo2021deep}
Iv{\'a}n L{\'o}pez-Espejo, Zheng-Hua Tan, John~HL Hansen, and Jesper Jensen. 2021.
\newblock Deep spoken keyword spotting: An overview.
\newblock \emph{IEEE Access}, 10:4169--4199.

\bibitem[{Lopez-Paz and Ranzato(2017)}]{lopez2017gradient}
David Lopez-Paz and Marc'Aurelio Ranzato. 2017.
\newblock Gradient episodic memory for continual learning.
\newblock \emph{Advances in neural information processing systems}, 30.

\bibitem[{Mandal et~al.(2014)Mandal, Prasanna~Kumar, and Mitra}]{mandal2014recent}
Anupam Mandal, KR~Prasanna~Kumar, and Pabitra Mitra. 2014.
\newblock Recent developments in spoken term detection: a survey.
\newblock \emph{International Journal of Speech Technology}, 17:183--198.

\bibitem[{Mazumder et~al.(2021)Mazumder, Banbury, Meyer, Warden, and Reddi}]{2021Few}
Mark Mazumder, Colby Banbury, Josh Meyer, Pete Warden, and Vijay~Janapa Reddi. 2021.
\newblock Few-shot keyword spotting in any language.
\newblock In \emph{Proc. Interspeech}.

\bibitem[{McCloskey and Cohen(1989)}]{mccloskey1989catastrophic}
Michael McCloskey and Neal~J Cohen. 1989.
\newblock Catastrophic interference in connectionist networks: The sequential learning problem.
\newblock In \emph{Psychology of learning and motivation}, volume~24, pages 109--165. Elsevier.

\bibitem[{Ng et~al.(2023)Ng, Xiao, Yip, Yang, Tian, Fu, Chng, and Ma}]{ng23b_interspeech}
Dianwen Ng, Yang Xiao, Jia~Qi Yip, Zhao Yang, Biao Tian, Qiang Fu, Eng~Siong Chng, and Bin Ma. 2023.
\newblock Small footprint multi-channel network for keyword spotting with centroid based awareness.
\newblock In \emph{Proc. Interspeech}, pages 296--300.

\bibitem[{Parisi et~al.(2019)Parisi, Kemker, Part, Kanan, and Wermter}]{cl}
German~I Parisi, Ronald Kemker, Jose~L Part, Christopher Kanan, and Stefan Wermter. 2019.
\newblock {Continual Lifelong Learning with Neural Networks: A Review}.
\newblock \emph{Neural networks}, 113:54--71.

\bibitem[{Parnami and Lee(2022)}]{parnami2022few}
Archit Parnami and Minwoo Lee. 2022.
\newblock Few-shot keyword spotting with prototypical networks.
\newblock In \emph{Proc. International Conference on Machine Learning Technologies (ICMLT)}, pages 277--283.

\bibitem[{Pelosin et~al.(2022)Pelosin, Jha, Torsello, Raducanu, and van~de Weijer}]{efcl1}
Francesco Pelosin, Saurav Jha, Andrea Torsello, Bogdan Raducanu, and Joost van~de Weijer. 2022.
\newblock Towards exemplar-free continual learning in vision transformers: an account of attention, functional and weight regularization.
\newblock In \emph{Proceedings of the IEEE/CVF Conference on Computer Vision and Pattern Recognition}, pages 3820--3829.

\bibitem[{Peng and Xiao(2024)}]{peng2024dark}
Tianyi Peng and Yang Xiao. 2024.
\newblock Dark experience for incremental keyword spotting.
\newblock \emph{arXiv preprint:2409.08153}.

\bibitem[{Petit et~al.(2023)Petit, Popescu, Schindler, Picard, and Delezoide}]{efcl2}
Gr{\'e}goire Petit, Adrian Popescu, Hugo Schindler, David Picard, and Bertrand Delezoide. 2023.
\newblock Fetril: Feature translation for exemplar-free class-incremental learning.
\newblock In \emph{Proceedings of the IEEE/CVF winter conference on applications of computer vision}, pages 3911--3920.

\bibitem[{Rebuffi et~al.(2017)Rebuffi, Kolesnikov, Sperl, and Lampert}]{icarl}
Sylvestre-Alvise Rebuffi, Alexander Kolesnikov, Georg Sperl, and Christoph~H Lampert. 2017.
\newblock {icarl: Incremental Classifier and Representation Learning}.
\newblock In \emph{Proc. the IEEE conference on Computer Vision and Pattern Recognition (CVPR)}, pages 2001--2010.

\bibitem[{Ritter et~al.(2018)Ritter, Botev, and Barber}]{ewc2}
Hippolyt Ritter, Aleksandar Botev, and David Barber. 2018.
\newblock Online structured laplace approximations for overcoming catastrophic forgetting.
\newblock \emph{Advances in Neural Information Processing Systems}, 31.

\bibitem[{Song et~al.(2024)Song, Liu, Yang, Peng, and Li}]{sc100}
Zeyang Song, Qianhui Liu, Qu~Yang, Yizhou Peng, and Haizhou Li. 2024.
\newblock Ed-skws: Early-decision spiking neural networks for rapid, and energy-efficient keyword spotting.
\newblock \emph{arXiv preprint arXiv:2406.12726}.

\bibitem[{Tang and Lin(2018)}]{tang2018kws}
Raphael Tang and Jimmy Lin. 2018.
\newblock Deep residual learning for small-footprint keyword spotting.
\newblock In \emph{2018 IEEE International Conference on Acoustics, Speech and Signal Processing (ICASSP)}, pages 5484--5488. IEEE.

\bibitem[{Wu et~al.(2019)Wu, Chen, Wang, Ye, Liu, Guo, and Fu}]{bic}
Yue Wu, Yinpeng Chen, Lijuan Wang, Yuancheng Ye, Zicheng Liu, Yandong Guo, and Yun Fu. 2019.
\newblock Large scale incremental learning.
\newblock In \emph{Proceedings of the IEEE/CVF conference on computer vision and pattern recognition}, pages 374--382.

\bibitem[{Xiao and Das(2024{\natexlab{a}})}]{cdoa}
Yang Xiao and Rohan~Kumar Das. 2024{\natexlab{a}}.
\newblock {Configurable DOA Estimation using Incremental Learning}.
\newblock \emph{arXiv preprint:2407.03661}.

\bibitem[{Xiao and Das(2024{\natexlab{b}})}]{ucil}
Yang Xiao and Rohan~Kumar Das. 2024{\natexlab{b}}.
\newblock {UCIL: An Unsupervised Class Incremental Learning Approach for Sound Event Detection}.
\newblock \emph{arXiv:2407.03657}.

\bibitem[{Xiao et~al.(2022{\natexlab{a}})Xiao, Hou, and Chng}]{cl2}
Yang Xiao, Nana Hou, and Eng~Siong Chng. 2022{\natexlab{a}}.
\newblock \href {https://doi.org/10.21437/Interspeech.2022-10500} {{Rainbow Keywords: Efficient Incremental Learning for Online Spoken Keyword Spotting}}.
\newblock In \emph{Proc. Interspeech}, pages 3764--3768.

\bibitem[{Xiao et~al.(2022{\natexlab{b}})Xiao, Liu, King, Singh, Chng, Plumbley, and Wang}]{cl3}
Yang Xiao, Xubo Liu, James King, Arshdeep Singh, Eng~Siong Chng, Mark~D Plumbley, and Wenwu Wang. 2022{\natexlab{b}}.
\newblock {Continual Learning for On-device Environmental Sound Classification}.
\newblock In \emph{Proc. the Detection and Classification of Acoustic Scenes and Events Workshop (DCASE)}.

\bibitem[{Yang et~al.(2022{\natexlab{a}})Yang, Lane, and Watanabe}]{yang2022online}
Muqiao Yang, Ian Lane, and Shinji Watanabe. 2022{\natexlab{a}}.
\newblock Online continual learning of end-to-end speech recognition models.
\newblock \emph{Proceedings of Interspeech}.

\bibitem[{Yang et~al.(2022{\natexlab{b}})Yang, Kim, Chung, and Chang}]{yang2022personalized}
Seunghan Yang, Byeonggeun Kim, Inseop Chung, and Simyung Chang. 2022{\natexlab{b}}.
\newblock Personalized keyword spotting through multi-task learning.
\newblock \emph{arXiv preprint arXiv:2206.13708}.

\bibitem[{Zhang et~al.(2018)Zhang, Suda, Lai, and Chandra}]{zhang2018hello}
Yundong Zhang, Naveen Suda, Liangzhen Lai, and Vikas Chandra. 2018.
\newblock Hello edge: Keyword spotting on microcontrollers.
\newblock \emph{arXiv:1711.07128}.

\bibitem[{Zhuang et~al.(2021)Zhuang, Lin, and Toh}]{brmp}
Huiping Zhuang, Zhiping Lin, and Kar-Ann Toh. 2021.
\newblock Blockwise recursive moore--penrose inverse for network learning.
\newblock \emph{IEEE Transactions on Systems, Man, and Cybernetics: Systems}, 52(5):3237--3250.

\bibitem[{Zhuang et~al.(2022)Zhuang, Weng, Wei, Xie, Toh, and Lin}]{zhuang2022acil}
Huiping Zhuang, Zhenyu Weng, Hongxin Wei, Renchunzi Xie, Kar-Ann Toh, and Zhiping Lin. 2022.
\newblock Acil: Analytic class-incremental learning with absolute memorization and privacy protection.
\newblock \emph{Advances in Neural Information Processing Systems}, 35:11602--11614.

\end{thebibliography}

\clearpage

\appendix

\begin{table*}[t!]
\centering
\caption{Overview of the three datasets used in our incremental KWS experiments.}
\resizebox{0.8\textwidth}{!}{
\begin{tabular}{lcccc} 
    \toprule
    \textbf{Dataset} & \textbf{Classes} & \textbf{Samples} & \textbf{Keyword Examples} & \textbf{Audio Duration} \\
    \midrule
    \textbf{GSC-V1} & 30 & 64,727 & \makecell{"yes", "no","up", "down"} & 1 sec each \\
    \textbf{GSC-V2} & 35 & 105,829 & \makecell{"yes", "no", "backward", "forward"} & 1 sec each \\
    \textbf{SC-100} & 100 & 313,951 & \makecell{"change", "turn",  "light", "door"} & 1 sec each \\
    \bottomrule
\end{tabular}}

\label{tab:dataset_overview}
\end{table*}

\section{Proof of equations}

\begin{proof}
    We first solve the recursive formulation for the \(\mathbb{A}_{t}\), the acoustic feature autocorrelation matrix (AFAM) from the task \(\tau_{t}\). According to the Woodbury
matrix identity, for any invertible square matrices we have \(A\) and \(C\), we have

\begin{small}
\begin{equation*}
    (A+UCV)^{-1} = A^{-1}-A^{-1}U(C^{-1}+VA^{-1}U)VA^{-1}.
\end{equation*}
\end{small}


\noindent Let \(A=\mathbb{A}^{-1}_{t-1}\), \(U=\mathbf{S}_t'{^\top}\), \(V=\mathbf{S}_t'\), and \(C=I\). Hence, from \(\mathbb{A}_{t} = (\mathbb{A}^{-1}_{t-1} + \mathbf{S}_t'{^\top}\mathbf{S}_t')^{-1}\) and the Woodbury
matrix identity, we have

\begin{small}
    \begin{equation*}
    \mathbb{A}_{t} = \mathbb{A}_{t-1} - \mathbb{A}_{t-1} \mathbf{S}_t'{^\top} (I + \mathbf{S}_t' \mathbb{A}_{t-1} \mathbf{S}_t'{^\top})^{-1} \mathbf{S}_t' \mathbb{A}_{t-1}
    \tag{a}
    \label{a}
\end{equation*}
\end{small}

\noindent which completes the proof for the recursive formulation of AFAuM. Now we proof calculate \(\widehat{{\theta}}_{\text{cls}}^{(t)}\).

\noindent Let \( Q_{t-1} = [\mathbf{S}_{0}'{^\top} \textbf{y}_0, \dots, \mathbf{S}_{t-1}'{^\top} \textbf{y}_{t-1}] \). According to~\eqref{eq7},~\eqref{eq6}, and~\eqref{a}, we have

\begin{align*}
    \widehat{{\theta}}_{\text{cls}}^{(t)} &= \mathbb{A}_{t} 
    \begin{bmatrix}
        Q_{t-1} - \mathbf{S}_t'{^\top} Y_t^{\text{train}}
    \end{bmatrix} \\
    &= \big[ \mathbb{A}_{t} Q_{t-1} - \mathbb{A}_{t} \mathbf{S}_t'{^\top} Y_t^{\text{train}} \big]
    \tag{b}
    \label{b}
\end{align*}

\noindent where

\begin{small}
    \begin{align*}
    \mathbb{A}_{t} Q_{t-1} &= \mathbb{A}_{t-1} Q_{t-1} \\ &- \mathbb{A}_{t-1} \mathbf{S}_t'{^\top} (I + \mathbf{S}_t' \mathbb{A}_{t-1} \mathbf{S}_t'{^\top})^{-1} \mathbf{S}_t' \mathbb{A}_{t-1} Q_{t-1}.
\end{align*}
\end{small}

\noindent This simplifies to:

\begin{small}
    \begin{align*}
    \widehat{{\theta}}_{\text{cls}}^{(t)}   &= \widehat{{\theta}}_{\text{cls}}^{(t-1)} \\   &- \mathbb{A}_{t} \mathbf{S}_t'{^\top} (I + \mathbf{S}_t' \mathbb{A}_{t-1} \mathbf{S}_t'{^\top})^{-1} \mathbf{S}_t' \mathbb{A}_{t-1} Q_{t-1}.
    \tag{c}
    \label{c}
\end{align*}
\end{small}

\noindent Let \( K_t = (I + \mathbf{S}_t' \mathbb{A}_{t-1} \mathbf{S}_t'{^\top})^{-1} \). Since,

\begin{equation*}
    I = K_t K_t^{-1} = K_t (I + \mathbf{S}_t' \mathbb{A}_{t-1} \mathbf{S}_t'{^\top}),
\end{equation*}

\noindent then we have

\begin{equation*}
    K_t = I - K_t \mathbf{S}_t' \mathbb{A}_{t-1} \mathbf{S}_t'{^\top}.
\end{equation*}

\noindent Thus, substituting in equation~\eqref{c},

\begin{align*}
    \widehat{{\theta}}_{\text{cls}}^{(t)} &= \widehat{{\theta}}_{\text{cls}}^{(t-1)} - \mathbb{A}_{t} \mathbf{S}_t'{^\top} K_t \mathbf{S}_t' \mathbb{A}_{t-1} Q_{t-1} \\
    &= \widehat{{\theta}}_{\text{cls}}^{(t-1)} - (\mathbb{A}_{t} - \mathbb{A}_{t-1}) Q_{t-1} \\
    &= (\mathbb{A}_{t} - \mathbb{A}_{t-1}) Q_{t-1} \\
    &= (\mathbb{A}_{t} - \mathbb{A}_{t-1}) \mathbf{S}_t'{^\top}.
\end{align*}

\noindent This allows equation~\eqref{c} to be reduced to:

\begin{equation}
    \widehat{{\theta}}_{\text{cls}}^{(t)} = \widehat{{\theta}}_{\text{cls}}^{(t-1)} - \mathbb{A}_{t} \mathbf{S}_t'{^\top} \widehat{{\theta}}_{\text{cls}}^{(t-1)}.
    \tag{d}
    \label{d}
\end{equation}

\noindent Finally, we could complete the proof by substituting equation~\eqref{d} into~\eqref{b}.




            
\end{proof}

\section{Details of datasets}
This section summarizes the three datasets used in our incremental KWS experiments: \textbf{GSC-V1}, \textbf{GSC-V2}, and \textbf{SC-100}. We list their core attributes, such as the number of classes, total samples, and the data pre-processing differences in ensuring a uniform 1-second duration per clip. In \textbf{SC-100}, each actual keyword utterance lasts between 0.4 and 1 second, and zero-padding is used at the beginning or end of the sample. This design also includes precise timestamp annotations for keyword onset and offset, enabling more refined early-decision analysis. By contrast, \textbf{GSC-V1} and \textbf{GSC-V2} only use zero-padding or truncation at the end of the audio clip and do not provide temporal boundaries for keyword occurrence.

The three datasets differ in their number of classes, total samples, and recording procedures. Table~\ref{tab:dataset_overview} outlines their main specifications, including examples of keywords, data sources, and additional information on background noise or speaker diversity. Each dataset has a fixed length of one second per audio clip. However, \textbf{SC-100} preserves more granular structure for the actual keyword utterance, using random zero-padding to maintain a total length of one second. In contrast, \textbf{GSC-V1} and \textbf{GSC-V2} do not provide specific onset or offset timestamps, which can obscure where the keyword appears within the audio.

\begin{table}[ht]
\centering
\caption{Incremental task division for different datasets. The format follows 
(Initial Task Size \(+\) Incremental Steps \(\times\) Classes per Step), where the model first 
trains on the initial task size and then progressively learn additional classes in 
multiple incremental steps.}
\resizebox{\columnwidth}{!}{
\begin{tabular}{lc}
    \toprule
    \textbf{Dataset} & \textbf{Incremental Task Division} \\ 
    \midrule
    \textbf{GSC-V1 (30 classes)}  
    & \makecell{15 \(+\) (5 \(\times\) 3) \\ 10 \(+\) (10 \(\times\) 2) \\ 10 \(+\) (20 \(\times\) 1)} \\ 
    \midrule
    \textbf{GSC-V2 (35 classes)}  
    & \makecell{15 \(+\) (5 \(\times\) 4) \\ 15 \(+\) (10 \(\times\) 2) \\ 15 \(+\) (20 \(\times\) 1)} \\ 
    \midrule
    \textbf{SC-100 (100 classes)} 
    & \makecell{50 \(+\) (10 \(\times\) 5) \\ 50 \(+\) (25 \(\times\) 2) \\ 50 \(+\) (50 \(\times\) 1)} \\ 
    \bottomrule
\end{tabular}}
\label{tab:incremental_tasks}
\end{table}

\section{Baseline details}
To comprehensively evaluate our proposed method on incremental KWS tasks, we compare it against six representative baselines from the incremental learning field:

\noindent \textbf{EWC} \citep{ewc}.
EWC limits forgetting by selectively restricting changes to crucial model parameters. It computes the Fisher Information Matrix (FIM) to estimate parameter importance and adds a quadratic penalty to discourage large shifts in these weights.

\noindent \textbf{Rwalk} \citep{rwalk}.
Rwalk improves upon EWC by introducing a path integral-based approach to track parameter changes throughout training. Additionally, it replays a small subset of past data, boosting adaptability while retaining older knowledge.

\noindent \textbf{iCaRL} \citep{icarl}.
iCaRL stores selected “exemplar” samples in a fixed-size memory buffer and employs a Nearest Mean-of-Exemplars (NME) classifier. This method thus blends replay with knowledge distillation to address forgetting.

\noindent \textbf{BiC} \citep{bic}.
BiC tackles class imbalance by adding a bias correction layer after the final classifier. Following a two-stage training plan, it first uses knowledge distillation and memory replay, then adjusts bias using a small validation set.

\noindent \textbf{RK} \citep{cl2}.
RK targets online KWS scenarios with limited resources. It uses a diversity-aware sampler that selects uncertain samples for a memory buffer. Together with data augmentation and knowledge distillation, this design helps reduce forgetting on edge devices.

\noindent \textbf{DE-KWS} \citep{peng2024dark}.
DE-KWS integrates dark knowledge distillation into a rehearsal-based pipeline. Besides storing past examples, it also keeps a log of pre-softmax logits to replay “dark” knowledge. Sampling and updating these logits throughout training lead to smoother task transitions and better model adaptability.

\section{Supplementary Experiment Results}

\begin{table}[t!]
\centering
\caption{Comparison of ACC, BWT, and TT for different exemplar-based methods with a buffer of size 1000 in the SC-100 dataset. We also compare them with our proposed exemplar-free method AnalyticKWS.}
\label{tab:acc_bwt_time}
\resizebox{0.9\linewidth}{!}{
\begin{tabular}{lccc}
    \toprule
    \textbf{Method} & \textbf{T=11} & \textbf{T=26} & \textbf{T=51} \\
    \midrule
    \multicolumn{4}{c}{\cellcolor{gray!20}\textbf{ACC (\%)}} \\
    \midrule
    RK     & 77.27  & 74.18  & 72.37 \\
    Rwalk  & 84.61   & 83.95 & 84.49 \\
    DE-KWS  & 74.91  & 67.61  & 63.70 \\
    iCaRL  & 75.48  & 51.00  & 26.05 \\
    BiC    & 69.50  & 70.41  & 70.26 \\
   \cellcolor[HTML]{C4D5EB}AnalyticKWS-128 & \cellcolor[HTML]{C4D5EB}\uline{85.77}  & \cellcolor[HTML]{C4D5EB}\uline{85.66}  & \cellcolor[HTML]{C4D5EB}\uline{85.55} \\
   \cellcolor[HTML]{C4D5EB}AnalyticKWS-256 & \cellcolor[HTML]{C4D5EB}\textbf{87.99}  & \cellcolor[HTML]{C4D5EB}\textbf{87.85}  & \cellcolor[HTML]{C4D5EB}\textbf{87.63} \\
    \midrule
    \multicolumn{4}{c}{\cellcolor{gray!20}\textbf{BWT}} \\
    \midrule
    RK     & -0.046 & -0.024 & -0.014 \\
    Rwalk  & -0.033 & -0.014 & -0.007 \\
    DE-KWS  & -0.045 & -0.024 & -0.014 \\
    iCaRL  & -0.049 & -0.039 & -0.035 \\
    BiC    & -0.069 & -0.028 & -0.016 \\
    \cellcolor[HTML]{C4D5EB}AnalyticKWS-128 & \cellcolor[HTML]{C4D5EB}\uline{-0.021} & \cellcolor[HTML]{C4D5EB}\uline{-0.008} & \cellcolor[HTML]{C4D5EB}\uline{-0.004} \\
   \cellcolor[HTML]{C4D5EB}AnalyticKWS-256 & \cellcolor[HTML]{C4D5EB}\textbf{-0.017} & \cellcolor[HTML]{C4D5EB}\textbf{-0.007} & \cellcolor[HTML]{C4D5EB}\textbf{-0.003} \\
    \midrule
    \multicolumn{4}{c}{\cellcolor{gray!20}\textbf{TT (s)}} \\
    \midrule
    RK     & 1141.46 & 691.95  & 439.79 \\
    Rwalk  & 810.47 & 797.58  & 790.24 \\
    DE-KWS  & 515.03  & 389.21  & 333.37 \\
    iCaRL  & 419.16  & 238.35  & 174.44 \\
    BiC    & 434.41  & 343.11  & 341.61 \\
    \cellcolor[HTML]{C4D5EB}AnalyticKWS-128 & \cellcolor[HTML]{C4D5EB}\uline{15.35} & \cellcolor[HTML]{C4D5EB}\uline{6.77} & \cellcolor[HTML]{C4D5EB}\uline{5.82} \\
   \cellcolor[HTML]{C4D5EB}AnalyticKWS-256 & \cellcolor[HTML]{C4D5EB}\textbf{15.67} & \cellcolor[HTML]{C4D5EB}\textbf{8.64} & \cellcolor[HTML]{C4D5EB}\textbf{7.12} \\
    \bottomrule
\end{tabular} }
\end{table}

Table~\ref{tab:acc_bwt_time} compares ACC, BWT, and TT across various exemplar-based methods (with a 1000-sample buffer) and our proposed exemplar-free AnalyticKWS variants. As the number of tasks grows from 
\(T=11\) to \(T=51\), rehearsal-based approaches like RK, DE-KWS, and BiC exhibit noticeable drops in accuracy and increasingly negative BWT values. iCaRL also suffers a drastic decline, suggesting it struggles to retain knowledge under large increments. In contrast, both AnalyticKWS-128 and AnalyticKWS-256 sustain the highest ACC scores (up to 87.63\%) while showing minimal forgetting, indicated by their near-zero BWT. Moreover, they complete training in only a few seconds per task, vastly outperforming all baselines in TT. These findings highlight that our analytic, exemplar-free approach effectively mitigates catastrophic forgetting while cutting computational costs and meeting the needs of real-world, resource-constrained keyword spotting.
\end{document}